\documentclass[aps,pra,twocolumn,nofootinbib,groupedaddress]{revtex4-1}

\usepackage{amsfonts,amssymb,bm}
\usepackage{amsmath}
\usepackage{graphicx}
\usepackage[english]{babel}
\usepackage{graphicx, lettrine}
\usepackage{epsfig}
\usepackage{natbib}
\usepackage{mathrsfs}
\usepackage{color}
\usepackage{subcaption}
        
\begin{document}
\newcommand{\red}[1] {\textcolor{red}{#1}}
\newcommand{\ket}[1]{\left| #1 \right\rangle}
\newcommand{\bra}[1]{\left\langle #1 \right|}
\newcommand{\braket}[2]{\left\langle #1 | #2 \right\rangle}
\newcommand{\braopket}[3]{\bra{#1}#2\ket{#3}}
\newcommand{\proj}[1]{| #1\rangle\!\langle #1 |}
\newcommand{\expect}[1]{\left\langle#1\right\rangle}
\newcommand{\Tr}{\mathrm{Tr}}
\def\Id{1\!\mathrm{l}}
\newcommand{\cM}{\mathcal{M}}
\newcommand{\cR}{\mathcal{R}}
\newcommand{\cE}{\mathcal{E}}
\newcommand{\cL}{\mathcal{L}}
\newcommand{\cl}{l}
\newcommand{\cH}{\mathcal{H}}
\newcommand{\cU}{\mathcal{U}}
\newcommand{\cP}{\mathcal{P}}
\newcommand{\reals}{\mathbb{R}}
\newcommand{\grad}{\nabla}
\newcommand{\rhohat}{\hat{\rho}}
\newcommand{\rhoMLE}{\rhohat_\mathrm{MLE}}
\newcommand{\rhotomo}{\rhohat_\mathrm{tomo}}
\newcommand{\diff}{\mathrm{d}\!}
\newcommand{\pdiff}[2]{\frac{\partial #1}{\partial #2}}
\newcommand{\todo}[1]{\color{red}#1}
\def\FCW{1.0\columnwidth}
\def\HCW{0.55\columnwidth}
\def\TPW{0.33\textwidth}
\def\tred#1{\textcolor{red}{#1}}
\def\cred#1{\textcolor{red}{(#1)}}

\newcommand\blfootnote[1]{%
	\begingroup
	\renewcommand\thefootnote{}\footnote{#1}%
	\addtocounter{footnote}{-1}%
	\endgroup
}


\title{An On-chip Homodyne Detector for Measuring Quantum States and Generating Random Numbers}

\author{ Francesco Raffaelli\textsuperscript{$\ddagger$}}
\author{Giacomo Ferranti\textsuperscript{$\ddagger$}}
\author{Dylan H. Mahler}
\author{Philip Sibson}
\author{Jake E. Kennard}
\author{Alberto Santamato}
\author{Gary Sinclair}
\author{Damien Bonneau}
\author{Mark G. Thompson}
\author{Jonathan C. F. Matthews}
\email[]{Jonathan.Matthews@Bristol.ac.uk}
\affiliation{Quantum Engineering Technology Labs, H. H. Wills Physics Laboratory and Department of Electrical \& Electronic Engineering, University of Bristol, BS8 1FD, UK.}

\blfootnote{\textsuperscript{$\ddagger$} These authors contributed equally to this work.}

\date{\today}

\begin{abstract}
Optical homodyne detection has found use in a range of quantum technologies as both a characterisation tool and as a way to post-selectively generate non-linearities. So far optical implementations have been limited to bulk optics.  Here we present the first homodyne detector fully integrated with silicon photonics and suitable for measurements of the quantum state of the electromagnetic field.  This high speed, compact detector shows low noise operation, with 10 dB of clearance between shot noise and electronic noise, up to a speed of 160 MHz. These performances are suitable for on-chip characterisation of optical quantum states, such as Fock or squeezed states. As a first application, we show the generation of quantum random numbers at 1.2~Gbps generation rate. The produced random numbers pass all the statistical tests provided by the NIST statistical test suite.
\end{abstract}

\maketitle
Homodyne detectors are ubiquitous across quantum optics.  They are used to measure quantum states~\cite{vogel89,breitenbach97,Neergard-Nielsen2007,15dBsqueezing} and characterise quantum processes~\cite{cspt1,Kumar2013}.  They find applications in continuous variables quantum computation and quantum key distribution~\cite{CV_review} and they enable sub-shot-noise quantum interferometry~\cite{Schnabel:2010}. But the interferometric stability required for both the creation of non-classical states of light and for subsequent homodyne detection is limiting even in small-scale experiments, requiring active stabilisation to compensate. To address this, we present a homodyne detector with all the necessary photonic components integrated onto a silicon chip.

Integrated quantum photonics \citep{jer1} is an approach aimed at miniaturising quantum optics components onto monolithic components in an effort to increase the scale with which phase stable quantum optics can be implemented. This includes reconfigurable nested waveguide interferometry, on-chip optical nonlinearity and on-chip detectors~\cite{jer3}.
Most recently, cryogenically cooled superconducting nanowire single photon detectors (SNSPD) have been integrated with electrically driven sources of single photons~\cite{Khasminskaya:2016aa}. But to date, more general quantum states of light that are generated~\cite{on_chip_squeezing} or manipulated~\cite{on_chip_entanglement} on-chip are still characterised off-chip, after undergoing a significant amount of coupling loss. By monolithic CMOS-compatible fabrication of homodyne detectors in silicon photonics, we aim to open up the prospect of measuring and fully characterising the quantum optics being explored and developed on-chip~\cite{jer3}.

The optical components required for one homodyne detector are a phase shifter, a balanced two-mode optical beamsplitter and two photodiodes. In the silicon-on-insulator (SOI) architecture, each of these components operate at room temperature and the required integrated photonics are commercially available from foundries. Integrated balanced detectors made of the same components have application in classical photonics~\cite{Hai2013,Cox2014}. However, the full potential of the homodyne detector lies with its ability to detect extremely weak fields---even down to the single photon level---by measuring the field's interference with a bright laser, that acts as a local oscillator (LO), at an optical beamsplitter. Ideally, measurement of the difference in the photocurrents in the two photodiodes is proportional to the quantum quadrature operator
\begin{equation}\label{quad}
\hat{Q}(\phi) = \hat{a} e^{i\phi}+\hat{a}^\dagger e^{-i \phi}
\end{equation}
where $\phi$ is the optical phase difference between the LO and the signal field and the operators $\hat{a}$ and $\hat{a}^\dagger$ are the lowering and raising operators of the electromagnetic field. From measurements of $\hat{Q}(\phi)$ for different local oscillator phases, it is possible to reconstruct the quantum state of the signal field in the optical mode that is given by the local oscillator. This process of measurement and reconstruction is called optical homodyne tomography and has been studied in great detail~\cite{lvovsky_review}. 

When the quantum electromagnetic vacuum field is incident onto the detector, the fluctuations in measurement outcomes can be exploited to generate random numbers~\cite{Gabriel2010}.  This is useful because random numbers find applications in many different fields, including cryptography, computational simulation and fundamental science, but true randomness cannot be generated with a classical computer because these so-called pseudo-random numbers generated with software can in-principle be predicted. In contrast with pseudo random number generators, quantum random number generators (QRNGs) rely on the outcomes of inherently non-deterministic quantum processes to generate random numbers that cannot be predicted \citep{high_bitrate_qrng,68gbps_qrng,ultrafast_qrng,Gabriel2010,qrng}. Examples of compact QRNGs have been recently demonstrated \cite{InP_qrng,phone_qrng}.

\begin{figure}[htbp!]
\centering
\includegraphics[width=\columnwidth]{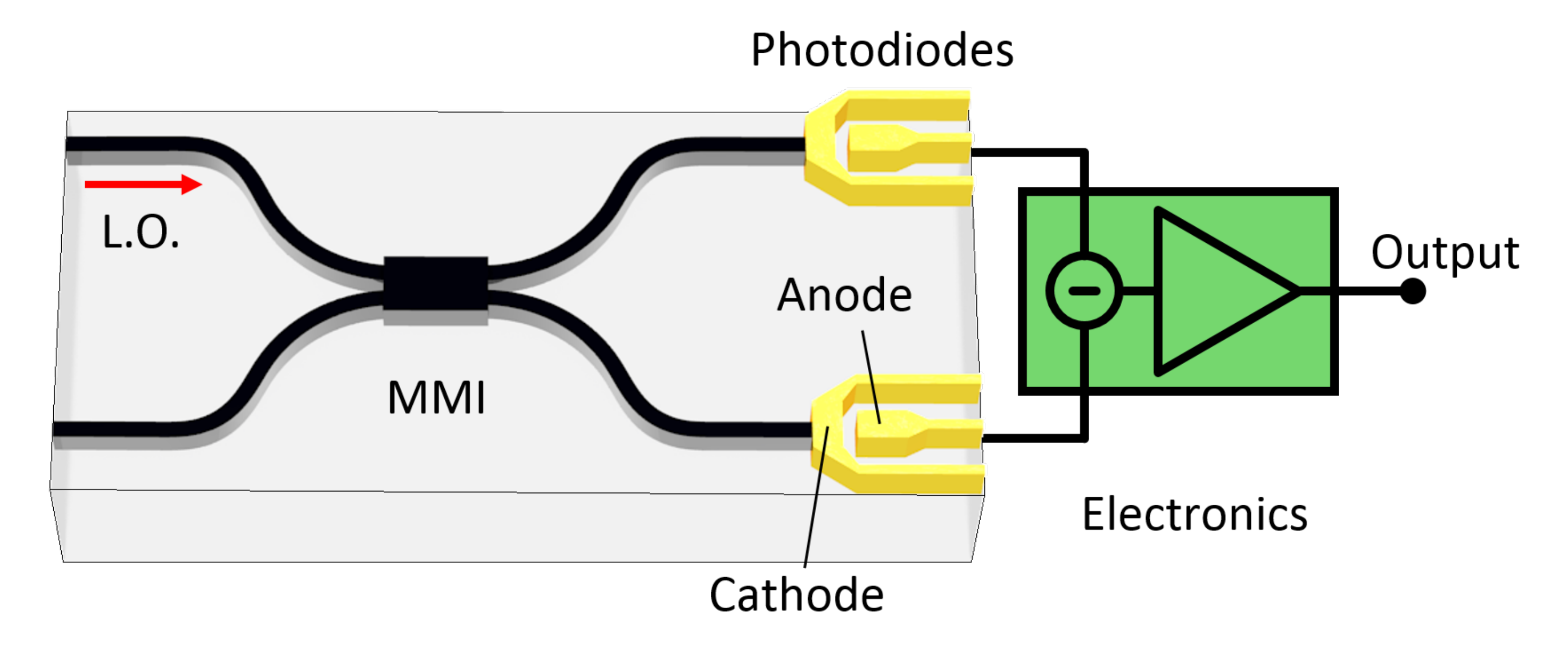}
\captionsetup{justification=raggedright, singlelinecheck=false}	
\caption{\textit{\textbf{Schematic of the device.}  The LO and the optical signal field are confined in waveguides. In our demonstration, the signal field is on a vacuum state. The beam-splitting operation is performed by a multi-mode interferometer (MMI). The two outputs of the MMI are coupled into two on-chip Ge photodiodes, generating two currents that are subtracted from each other by electronics.}}
\label{chip}
\end{figure}

The photonics for the homodyne detector reported here (Fig.~~\ref{chip}) were fabricated on a SOI chip as part of a multi-project wafer run organised by IMEC foundry services. The LO was an external laser source and it was coupled into single-mode waveguide within the chip via a vertically coupled grating. We used a multimode interference device (MMI) as the beam-splitter, with each output coupled via waveguide to a reverse-biased on-chip Germanium photodiode. The photo-currents generated by the two photodiodes were then subtracted from each other and amplified by a transimpedance amplifier---our electronics  configuration was based on the design in Ref.~\onlinecite{Kumar20125259}. The entire system, inclusive of the silicon chip and the PCB for the electronics, is a few centimetres square and the total footprint of photonics is $<1$~mm$^2$. 

Ambient noise (optical and electronic), dark current from the photodiodes, and experimental instability all contribute to noise in the quadrature measurement. These manifestations of noise can be modelled as optical loss in the channel of the signal field \cite{Appel2007}, which together with optical loss in the beamsplitter and the individual efficiency of the photodiodes, defines the overall efficiency $\eta$ of the detector. In our device, we identified three sources of inefficiency: the electronic noise generated by the detection circuit, the optical loss in the MMI and the inefficiency of the photodiodes. We measured the photodiodes effective responsivity as a single system, obtaining a value of 0.8~A/W in both arms, corresponding to an estimated quantum efficiency of $\eta_{pd}$=0.64.

\begin{figure}[htbp!]
	\centering
	\includegraphics[width=.72\columnwidth]{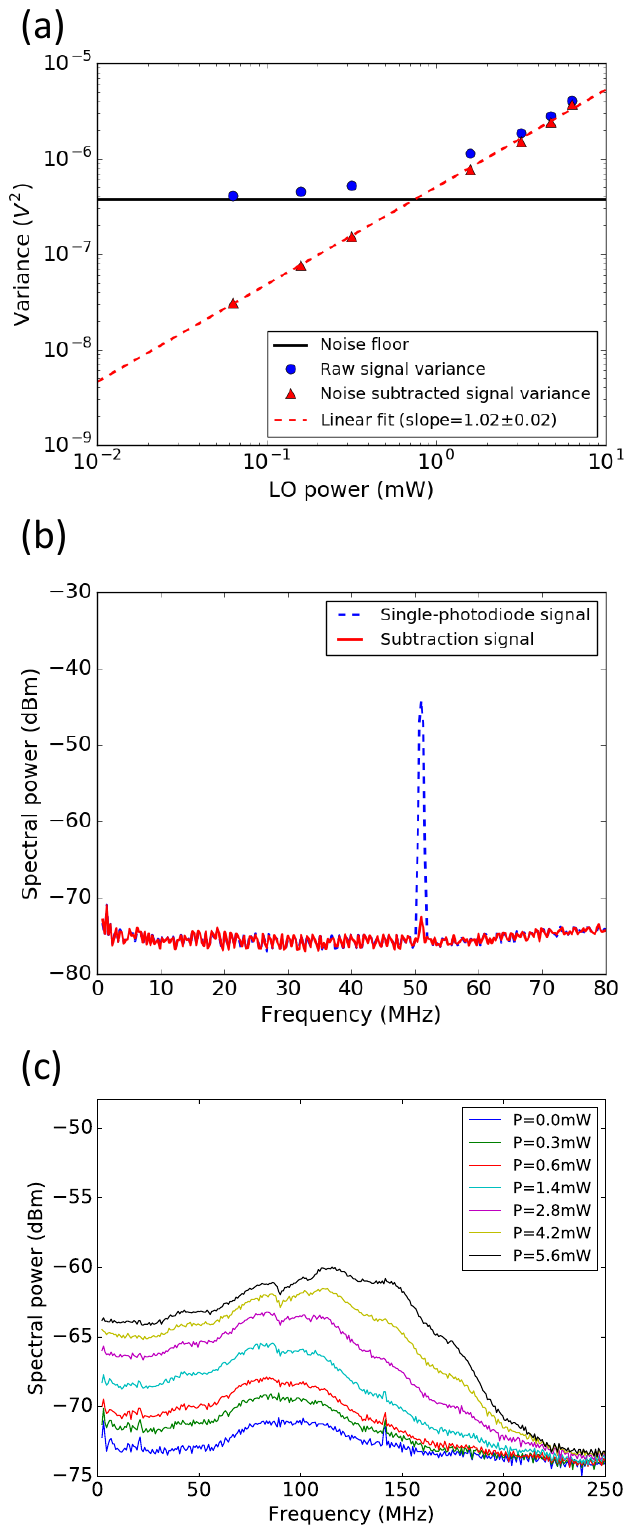} 
	\captionsetup{justification=raggedright, singlelinecheck=false}
	\caption{\textit{\textbf{Performance of the on-chip homodyne detector.}(a):Signal variance for different LO powers. The blue dots represent the raw signal variances, the red ones correspond to the noise-subtracted variances and the black line marks the variance of the electronic noise. The red line is a linear fit of the noise-subtracted variances with slope is $1.02~\pm~0.02$. (b): Measurement of the CMRR. Red line: Power spectrum of the signal with $18~\mu{W}$ of pulsed LO provided by a Pritel FFL-50 laser. Blue line: Power spectrum of the signal with the same LO, but with one photodiode disconnected from the circuit. The difference in height between the peaks is a measurement of the CMRR.
			(c): Spectral response of the integrated homodyne detector for different LO powers. The graph shows a SNC of 10~dB for a LO power of 5.6~mW over a bandwidth of $\sim$160 MHz. These values have been measured using a CW LO at a wavelength of 1550~nm.}}
	\label{performancev2}
\end{figure}

\begin{figure}[b!]
	\centering
	\includegraphics[width=0.9\columnwidth]{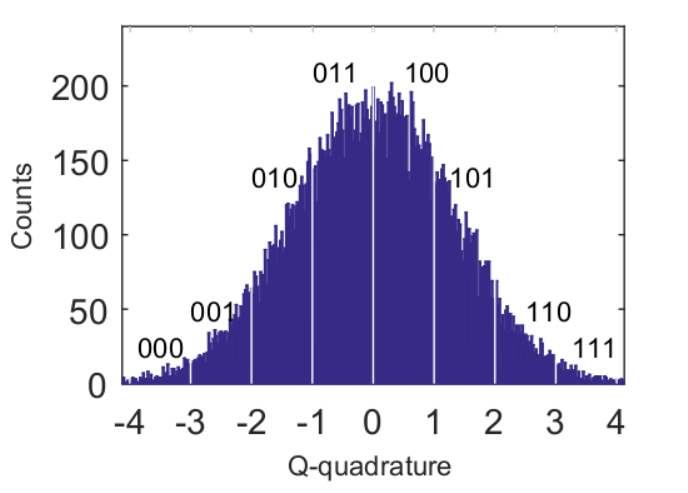}\hspace{3ex}
	\captionsetup{justification=raggedright, singlelinecheck=false}	
	\caption{\textit{\textbf{Measured histogram of the shot-noise signal.} The quadratures have a Gaussian distribution. The corresponding shot-noise histogram is divided into $2^n$ bins and each bin is labelled with a n-bit string which is used to label 		each sample from the oscilloscope. Since the outcomes are unpredictable, a bit string composed of all the samples will be random. We illustrate with n=3 bits as an example.}}
	\label{protocol}
\end{figure}

The electronic noise is a gaussian-distributed random quantity which can be measured directly in the absence of a LO. With an optical signal present, the electronic output will be gaussian-distributed, with a variance given by the sum of the variances of electrical signal and noise. So the variance of the noise-free signal can be estimated from: 
\begin{equation}
	\sigma_{SN}^2=\sigma_{O}^2-\sigma_{EN}^2,
	\label{sigma_quant}
\end{equation}
where $\sigma_{O}$ is the standard deviation of the raw output of the detector, $\sigma_{SN}$ is the standard deviation of the shot-noise contribution --- the fundamental quantum noise of the light field --- and $\sigma_{EN}$ is the standard deviation of the electronic technical noise contribution.

The ratio between the variance of the raw output of the detector measured at the highest LO power (6.3~mW) and zero LO power is $\sim$10~dB (see Fig.~~\ref{performancev2}a). The plot in Fig.~\ref{performancev2}a) also shows that the noise-subtracted variances on a bi-logarithmic scale are very well fitted by a line of slope 1.02$\pm$0.02, confirming the linear dependence on LO power, which agrees with the expected manifestation of quantum vacuum fluctuations as gaussian-distributed white noise.
The corresponding efficiency of the homodyne detector is given by~\cite{Kumar20125259}:
$$
\eta_{SNR}=1-\frac{\sigma_{EN}^2}{\sigma_{O}^2}=0.9
$$
which, combined with the photodiodes contribution, leads to a total detector efficiency of $$\eta=\eta_{pd}*\eta_{SNR}=0.58.$$ 
This value is high enough to characterise the quantum features of optical states \cite{low_eff_fringes,low_eff_tomo}.  

The common mode rejection ratio (CMRR) of a homodyne detector is defined as the ratio between the signal measured when only one of the photodiodes is illuminated and when both are. This quantifies how well the balanced signals from the two photodiodes can suppress technical noise in the LO. The graph in Fig.~\ref{performancev2}b) shows the spectra that we measured with our device, of both subtracted and unsubtracted signals measured using a pulsed LO with a repetition rate of 50 MHz. The difference between the heights of the two peaks at 50 MHz corresponds to the CMRR which we found to be 28 dB. This value is already  sufficient for continuous variables quantum information --- similar CMRRs have been used to characterise squeezed states~\cite{Ast13}. 

The bandwidth of a homodyne detector defines the speed with which it can be maximally operated. It defines the maximum spectral width that the signal field can have in order to be measured efficiently. The measured spectral response of our detector is shown in Fig.~\ref{performancev2}c) and the point at which it decreases to 3~dB below its peak value is $\sim$160~MHz. Since integrated photodiodes have a very low capacitance, the bandwidth of our detector is one and a half times higher than a comparable detector design implemented with conventional bulk optics \cite{Kumar20125259}.

The quadrature measurements $\hat{Q}$ for the vacuum states are non-deterministic and follow a Gaussian probability distribution 
\begin{equation}
	P(\hat{Q})=\frac{1}{\sqrt{\pi}}e^{\frac{-\hat{Q}^2}{\hbar}},
	\label{marginal_distribution}
\end{equation}
as shown in Fig.~\ref{protocol}. To extract random numbers, the range of possible measurement outcomes were divided into $2^n$ equally spaces bins and each bin is labelled with an $n-$bit string,  (Fig.~\ref{protocol}). Thus each measurement outcome corresponded to the generation of an $n-$bit number.

To be compatible with randomness extraction hardware, we used equally spaced bins, but this means the bits strings associated with the central bins will be more likely to appear, skewing the randomness of the random bits. Moreover correlations in the electronic background noise could be used by an adversary. We therefore implemented the Toeplitz hashing algorithm~\cite{post_processing_qrng} as a randomness extractor with a desktop computer. We calculated the min-entropy which describes the amount of extractable randomness from the quantum signal distribution. It is defined as 
\begin{equation}
\mathrm{H}_{\infty}=-\mathrm{log_2}( \underset{x\in \{0,1\}^{n}}{\mathrm{max}}{\mathrm{Pr}[X=x]}),
\label{min_entropy}
\end{equation}
where $X$ corresponds to the quantum signal shot-noise distribution over $2^n$ bins, and $Pr[X=x]$ is the probability to obtain a particular value for $X$.  In homodyne detection, and in fact in most QRNG schemes, we do not have direct information about the quantum signal distribution because it is always mixed with some classical noise. We estimated the true quantum variance, under the assumption of a Gaussian distribution using Eqn. \ref{sigma_quant}. The calculated min-entropy was $\sim$5.92~bits/sample when sampling the raw data at 8 bits/sample. Fig.~\ref{performancev2}c) shows that $\sigma_{O}$ is at least 3 dB above $\sigma_{EN}$ up to 200~MHz.  This implies that sampling at a rate of 200~Msamples/sec will not introduce additional correlations in the sampled bits.  Further justification of this sampling rate can be found in the Supplementary Information. At this sampling rate, quantum random numbers are generated at a rate of $\sim1.2$~Gbps. We then tested the generated random bits with the statistical tests provided in Ref.~\onlinecite{nist_web}. Our QRNG passed all the tests for unbiased random numbers from the NIST statistical test suite~\cite{nist_web}, as reported in table \ref{nist_table}.

\begin{table}[ht!]
\centering 
\setlength{\tabcolsep}{8pt}
\begin{tabular}{|l|c|} 
\hline

 Test name &  Success rate   \\ [0.5ex]
\hline 
\hline
Frequency                  &  0.996       \\
Block Frequency            &  0.998       \\
Cumulative Sums            &  0.994       \\ 
Runs                       &  0.990       \\
Longest Run                &  0.990       \\
Rank                       &  0.990       \\ 
FFT                        &  0.987       \\
Non Overlapping Template   &  0.990       \\
Overlapping Template       &  0.991       \\
Universal                  &  0.992       \\
Approximate Entropy        &  0.987       \\
Random Excursions          &  0.993       \\
Random Excursions Variant  &  0.995       \\
Serial                     &  0.989       \\
Linear Complexity          &  0.989       \\
 [1ex] 
\hline
\end{tabular}

\captionsetup{justification=raggedright,singlelinecheck=false}	
\caption{\textit{ \textbf{Statistical tests on the random data.}\\ We report the results for NIST statistical test suite. The set of data, composed of $10^9$ bits, was divided into 1000 blocks and the randomness tests were applied to each block. Standard practice~\cite{nist_web} dictates that a success rate above 0.98 was required to constitute a pass. Therefore we observe that the generated data passed all the randomness tests provided.}}  
\label{nist_table} 
\end{table}

The integrated homodyne detector that we have designed, implemented and characterised is fast, low noise and compact. This is key to developing fully integrated quantum photonics in the continuous variables regime of quantum optics. The compact design is compatible with complex and reconfigurable interferometry~\cite{on_chip_entanglement} and the lithographic manufacture is amendable to potentially high-yield enabling many-mode quantum characterisation using multiple homodyne detectors implemented on one chip.  
We anticipate application of integrated homodyne detectors in quantum cryptography components \citep{on_chip_qkd} and on-chip quantum sensing with squeezed states of light.

\begin{acknowledgments}
We thank Magnus Loutit for technical assistance. This work was supported by ERC, PICQUE, BBOI,  QUCHIP, the €‹US Army Research Office (ARO) Grant No. W911NF-14-1-0133, the
EPSRC Quantum Communications Hub (EP/M013472/1) and the EPSRC programme grant (EP/L024020/1). ‹MGT and JCFM acknowledge fellowship support from the EPSRC.
\end{acknowledgments}

\section{Methods}\label{methods}

\subsection{Electronics}
The design of the detection electronics has been based on the one developed in \cite{Kumar20125259}. The subtraction signal generated by the two photodiodes is amplified by a OPA847 operational amplifier in transimpedance configuration. The voltages supplying it have been stabilised by means of two fixed voltage regulators (LM78L05 and LM79L05). The same circuit also provides the voltages reverse-biasing the photodiodes, which are stabilised by two adjustable voltage regulators (LM317LM and LM337LM). We chose these components to be adjustable in order to mantain some control over the time response of the photodiodes. 

\end{document}